\documentclass[conference]{IEEEtran}
\IEEEoverridecommandlockouts
\usepackage{lipsum} 
\usepackage{cite}
\usepackage{amsmath,amssymb,amsfonts}
\usepackage{subfigure, url, doi, xspace}
\usepackage{autobreak}
\usepackage{algorithm, algorithmic}
\usepackage{graphicx}
\usepackage{epsfig}
\usepackage{textcomp}
\usepackage{xcolor}
\usepackage{epstopdf}
\def\BibTeX{{\rm B\kern-.05em{\sc i\kern-.025em b}\kern-.08em
    T\kern-.1667em\lower.7ex\hbox{E}\kern-.125emX}}
    
\begin{document}

\title{Projected Multi-Agent Consensus Equilibrium for Ptychographic Image Reconstruction \\
\thanks{Zhai and Wohlberg were partially supported by Los Alamos National Laboratory's Directed Research and Development Program under project 20200061DR. Buzzard and Bouman were partially supported by the NSF under award  CCF-1763896.}
}
\author{
\IEEEauthorblockN{Qiuchen Zhai}
\IEEEauthorblockA{\textit{School of Electrical and Computer Engineering}\\
\textit{Purdue University}\\
West Lafayette, IN, USA \\
qzhai@purdue.edu} \\   
\IEEEauthorblockN{Gregery T. Buzzard}
\IEEEauthorblockA{\textit{Department of Mathematics} \\
\textit{Purdue University}\\
West Lafayette, IN, USA \\
buzzard@purdue.edu} \\
\and
\IEEEauthorblockN{Brendt Wohlberg}
\IEEEauthorblockA{\textit{Theoretical Division} \\
\textit{Los Alamos National Laboratory}\\
Los Alamos, NM, USA \\
brendt@ieee.org}  \\   
\IEEEauthorblockN{Charles A. Bouman}
\IEEEauthorblockA{
\textit{School of Electrical and Computer Engineering}\\
\textit{Purdue University}\\
West Lafayette, IN, USA \\
bouman@purdue.edu}
}

\maketitle
\begin{abstract} Ptychography is a computational imaging technique using multiple, overlapping, coherently illuminated snapshots to achieve nanometer resolution by solving a nonlinear phase-field recovery problem.  
Ptychography is vital for imaging of manufactured nanomaterials, but existing algorithms have computational shortcomings that limit large-scale application.  

In this paper, we present the Projected Multi-Agent Consensus Equilibrium (PMACE) approach for solving the ptychography inversion problem.  
This approach extends earlier work on MACE, which formulates an inversion problem as an equilibrium among multiple agents, each acting independently to update a full reconstruction.  
In PMACE, each agent acts on a portion (projection) corresponding to one of the snapshots, and these updates to projections are then combined to give an update to the full reconstruction.  
The resulting algorithm is easily parallelized, with convergence properties inherited from convergence results associated with MACE.  
We apply our method on simulated data and demonstrate that it outperforms competing algorithms in both reconstruction quality and convergence speed.

\end{abstract}

\begin{IEEEkeywords}
Ptychography, Consensus Equilibrium, Coherent Imaging, Phase Retrieval
\end{IEEEkeywords}

\section{Introduction}
\label{sec:intro}
Ptychography is a computational imaging technique in which a coherent scanning probe is moved across an object while recording the resulting far-field diffraction pattern \cite{pfeiffer2018x}. 
The probe is moved so that each illuminated region has substantial overlap with neighboring regions; this overlap provides redundant information that can be used to computationally retrieve the relative phase of the Fraunhofer diffraction plane. 
In this way the full complex transmittance image can be recovered from the intensity measurements, thus providing a detailed visualization of the object. 
Ptychographic methods can achieve resolution on the scale of a few nanometers, which makes them crucial for imaging manufactured nanomaterials.

A variety of numerical methods have been proposed for iterative phase retrieval from phaseless measurements. One class of methods uses alternating projections between a constraint set in the Fourier domain (to fit measured data) and a constraint set in the physical domain (to enforce nonnegativity or other properties). 
This class of methods includes error reduction (ER) \cite{fienup1982phase} and several variants, including hybrid input-output (HIO) \cite{fienup1982phase}, difference map (DM) \cite{elser2003phase}, averaged successive reflections (ASR) \cite{bauschke2002phase}, and relaxed averaged alternating reflections (RAAR) \cite{luke2004relaxed}. 
ER alternates projections between the two constraint sets to update the estimate, while HIO improves convergence by modifying the projection function in the Fourier domain. 
ASR can be interpreted as the Douglas–Rachford algorithm applied to phase retrieval problems with a nonconvex Fourier constraint. 
The RAAR algorithm further improves convergence with a relaxation strategy to combine the ASR algorithm with the projection operator in the Fourier domain. 
Scalable hetereogeneous adaptive real-time ptychography (SHARP) \cite{marchesini2016sharp} is a variant of the RAAR algorithm for ptychographic image reconstructions.
Though these alternating projection methods are parallelizable, they are not guaranteed to converge to an optimal solution \cite{marchesini2016alternating}.

Another class of algorithms derives from the ptychographical iterative engine (PIE) \cite{rodenburg2004phase}, as revised for serial ptychographic image reconstruction.
Algorithms related to this approach include PIE\cite{rodenburg2004phase}, extended PIE (ePIE) \cite{maiden2009improved}, regularized PIE (rPIE) and mPIE \cite{maiden2017further}.
In each iteration, the PIE-type algorithms process the intensity measurements one at a time to revise the estimates in a stochastic gradient approach \cite{maiden2017further}.
The PIE algorithm and its variants have fast convergence rate. 
However, the accelerated Wirtinger Flow (AWF) algorithm \cite{xu2018accelerated} has been shown to have faster convergence rate and is more robust to noise than earlier algorithms \cite{xu2018accelerated}.  

In this paper, we present the Projected Multi-Agent Consensus Equilibrium (PMACE) approach, which extends earlier work on the Multi-Agent Consensus Equilibrium (MACE) framework \cite{buzzard2018plug}.
The MACE framework formalizes and extends solutions found by the Plug-and-Play algorithm (PnP) \cite{venkatakrishnan2013plug, sreehari2016plug}. MACE formulates the inversion problem as a set of equilibrium equations that balance the effect of multiple agents, each with independent updates to a proposed solution. 
In PMACE, each agent acts independently on a patch of the complex transmittance, with each patch corresponding to the measurements from one probe position. The updates to the patches are then reconciled by a carefully chosen weighted average to update the full reconstruction.  

The resulting PMACE approach is easily parallelized and is guaranteed to converge under appropriate hypotheses since it inherits convergence results associated with the MACE framework.  
We compare our method with competing algorithms including AWF \cite{xu2018accelerated}, SHARP \cite{marchesini2016sharp}, and SHARP+, which is derived directly from RAAR \cite{luke2004relaxed} but which has a different update function than SHARP.
We apply our method on simulated noise-free and noisy data and demonstrate that it outperforms competing algorithms in both convergence rate and reconstruction quality.

\section{Problem Formulation}
\label{sec:problem formulation}
In ptychography, an object is illuminated by a coherent x-ray probe at various positions, and the exit wave from a given probe position is recorded by a detector at the far-field Fraunhofer diffraction plane. The goal is to recover the complex transmittance of the object from intensity measurements.  In this work, we assume the complex probe profile is known.  

Let $x \in \mathbb{C}^{N_{1} \times N_{2}}$ be the unknown complex transmittance, and let $d\in \mathbb{C}^{N_{p} \times N_{p}}$ be the known complex probe illumination. 
For each probe location indexed by $j \in \{ 0, \dots, J-1 \}$, let $z_{j}\in \mathbb{R}^{N_{p} \times N_{p}}$ be the measured data array recorded on a detector with $N_{p} \times N_{p}$ pixels. 

We describe an idealized forward model taking transmittance to measurements by first windowing $x$ to obtain a patch corresponding to one probe position, multiplying by the complex illumination, then taking the Fourier transform.  We combine this with a Poisson distribution to model the detector response as
\begin{equation}
\label{eq: func_forward}
z_{j} = \mathrm{Pois}(| \mathcal{F} D P_{j} x |^2) \;,
\end{equation}
where $\mathcal{F}$ denotes the 2D orthonormal Fourier transform matrix, $D = \mathrm{Diag}(d)$  is a diagonal matrix representing the complex illumination, and $P_{j}: \mathbb{C}^{N_{1} \times N_{2}} \rightarrow \mathbb{C}^{N_{p} \times N_{p}}$ is a projection that extracts one patch from the complex image. Here and below, the absolute value is applied pointwise.  

We convert to amplitude in a patch by defining $x_{j} = P_{j} x$ to be the patch corresponding to probe location indexed by $j$ and $y_{j}$ to be the square root of the measurement $z_{j}$.  This gives measured amplitudes
\begin{equation}
y_{j} = \sqrt{\mathrm{Pois}(| \mathcal{F} D x_{j} |^2) } \;.
\end{equation}

In our approach, the problem is solved using the maximum likelihood (ML) estimate, which is given by
\begin{equation}
x^* = \arg\min_{x}\left \{\sum_{j=0}^{J-1}f_{j}(x_{j}) \right \} \;,
\end{equation}
where $f_{j}(x_{j})$ is a cost function that enforces data fidelity.  Since the square root is an approximate variance-stabilizing transform for the Poisson distribution, we use squared error as a rough approximation of negative log-likelihood.  This yields
\begin{equation} \label{eq1}
\begin{split}
f_{j} \left ( x_{j} \right ) 
& =  \frac{1}{2\sigma_n^2} \left \| \, y_{j} - | \mathcal{F} D x_{j}| \, \right \|^2 \;,
\end{split}
\end{equation}
where $\sigma_n^2$ is an estimate of the noise variance in $y_j$.

\section{Reconstruction Algorithms}
\label{sec:reconstruction_algorithm}
In this section, we describe PMACE and SHARP+ for the ptychographic reconstruction problem. The PMACE approach extends the MACE framework, which formulates the inversion problem using a set of equilibrium equations.  SHARP+ is an extension of the RAAR algorithm to ptychography. 

\subsection{PMACE Approach}
To introduce the PMACE formulation for ptychographic image reconstruction, we begin with the agent in our approach. Each agent is an operator that updates the current estimate of a single projection.
In this work, we use probe-weighted proximal maps as agents. The standard proximal map for $f_j$ is 
\begin{equation}
    \begin{aligned}
        L_{j} (x_{j}) = \arg \min _ {v} \left \{ f_{j}(v) + \frac{1}{2 \sigma^2} \left \| v - x_{j} \right \|^2 \right \} \;,
    \end{aligned}
\end{equation}
which comes from the ADMM algorithm applied to minimize the sum of the $f_j$.   This proximal map is reinterpreted in Plug-and-Play \cite{sreehari2016plug} using a Bayesian framework, with $f_j$ as a data-fitting term and the squared norm as a prior term for a Gaussian distribution with mean $x_j$ and variance ${\sigma}^{2} / {\sigma}_{n}^{2}$. 

However, illumination by $d$ introduces uncertainty in the estimate of $x_j$, so we model the distribution of of $v - x_j$ by a zero-mean Gaussian with variance proportional to $|D|^{-2}$; we incorporate this by replacing $v - x_j$ with $Dv - D x_j$ inside the norm.  
Since the discrete Fourier transform operator $\mathcal{F}$ is orthonormal, we can also include it inside the norm to obtain
\begin{align} \label{eq:Fjv}
    F_{j} (x_{j})
        &= \arg \min _ {v} \left \{  f_{j}(v) + \frac{1}{2 \sigma^2} \left \| \mathcal{F} D v - \mathcal{F} D x_{j} \right \|^2 \right \}.
\end{align}
Introducing new variables $u = \mathcal{F} D v$ and $w_j = \mathcal{F} D x_j$, we can obtain  $F_j(x_j)$ from $u^* = \mathcal{F} D v^*$, where $v^*$ is the solution of \eqref{eq:Fjv} and we use \eqref{eq1} to rewrite $f_j$ to get
\begin{align} \label{eq:Fju}
    u^* = \arg \min_u \left \{ \frac{1}{2} \|y_j - |u| \|^2 + \frac{\sigma_n^2}{2 \sigma^2} \|u - w_j\|^2 \right \}.
\end{align}
Since the complex argument/phase of $u$ is not constrained by $y_j$, it must equal the phase of $w_j$, so we can write $u = r \odot \frac{w_j}{|w_j|}$, where $r$ is nonnegative, $\odot$ is the Hadamard product, and the fraction is 0 where $w_j$ is 0.
Using this in \eqref{eq:Fju}, we have 
\begin{align}
    r^* =  \arg \min_r \left \{ \frac{1}{2} \|r - y_j \|^2 + \frac{\sigma_n^2}{2 \sigma^2} \|r - |w_j|\|^2 \right \}.
\end{align}
Taking $\alpha = {\sigma_{n}^{2}}/{\sigma^2}$, which is a measure of noise-to-signal ratio, and using the 1st order optimality condition, we obtain $r^* = (\alpha |w_j| + y_j) / (\alpha + 1)$.  Unwinding the variable definitions (and omitting $\odot$ for brevity) gives
\begin{equation}
    \begin{split}  \label{eq:F-interpolate}
        F_{j} (x_{j})
        & = \frac{\alpha x_{j} + D^{-1}\mathcal{F}^{*} \left ( y_{j} \frac{\mathcal{F} D x_{j}}{|\mathcal{F} D x_{j}|} \right )}{1 + \alpha} \;.
    \end{split}
\end{equation}

When $\alpha=0$, the probe-weighted proximal map simply matches the current estimate with the corresponding measurement and returns the closest data-fitting point $F_{j} (x_{j}) = D^{-1}\mathcal{F}^{*} \left ( y_{j} \frac{\mathcal{F} D x_{j}}{|\mathcal{F} D x_{j}|} \right )$.
In the limit as $\alpha$ approaches $\infty$, the probe-weighted proximal map function returns the current estimate $x_j$. 
Hence the agent in \eqref{eq:F-interpolate} interpolates between the current estimate $x_j$ and the closest data-fitting point, the interpolation being controlled by varying the value of $\alpha$.

For the MACE formulation, we stack the individual projections to obtain a vector $\mathbf{x} = [x_{0}, x_{1}, \dots, x_{J-1}]^t$.
Then we define a stacked forward operator $\mathbf{F}$ and a consensus operator $\mathbf{G}$ that computes a weighted average of each component and reallocates the results as
\begin{equation}
    \mathbf{F} (\mathbf{x}) = \begin{pmatrix}
    F_{0} (x_{0}) \\ 
    \vdots \\ 
    F_{J-1} (x_{J-1}) 
    \end{pmatrix} \;
    \textrm{and} \; \;
    \mathbf{G} \left ( \mathbf{x} \right ) = \begin{pmatrix}
    \bar{x}_{0}\\ 
    \vdots \\ 
    \bar{x}_{J-1}
    \end{pmatrix}.  
\end{equation}
Here the component vectors in $\mathbf{G}$ are determined by weighting by $|D|^\kappa$, back-projecting to the full image, then doing a weighted average of these back-projections.  This is given by
\begin{equation} \label{eq:xjbar}
    \bar{x}_{j} = P_{j} \Lambda^{-1}  \sum_{i=0}^{J-1} P_{i}^{t} |D|^{\kappa} x_{i} \;,
\end{equation}
where ${\Lambda} = \sum_{j=0}^{J-1} P_{j}^{t} |D|^{\kappa}$ and $\kappa$ denotes the probe exponent parameter. Intuitively, the consensus operator projects the image patches associated with the scan locations back to the full-size image and normalizes this image with $\Lambda$, which uses the projections and probe weighting so that overlapping areas are averaged appropriately. 
The probe exponent parameter $\kappa$ can be used to tune the weighted average to match the uncertainty in $x_j$ introduced by measurement uncertainty and by the probe.   The choice of probe exponent parameter can improve algorithm performance.

To obtain the PMACE solution, we solve the equation
   $ \mathbf{F}(\mathbf{x}) = \mathbf{G}(\mathbf{x}) $.
As shown in \cite{buzzard2018plug}, the solution can be solved as the fixed point of the map 
\begin{equation}
     \mathbf{T} = (2 \mathbf{G} -\mathbf{I})(2 \mathbf{F} - \mathbf{I}) ,
\end{equation}
which can be computed via the Mann iteration
\begin{equation}
    \mathbf{x} \gets (1 - \rho) \mathbf{x} + \rho \mathbf{T} \mathbf{x} \;,
\end{equation}
where $\rho \in (0,1)$ affects the convergence rate but not the final result. These iterates are guaranteed to converge  to a fixed point if $\mathbf{T}$ is non-expansive and has a fixed point.   

Algorithm~\ref{algorithm_PMACE} shows pseudocode for computing the PMACE solution. 
The algorithm uses an initial $\mathbf{x}^{(0)} \in \mathbb R^{N_{1} \times N_{2}}$ to construct stacked vectors as above, then uses Mann iterations to update the estimates of projections.  The final reconstruction result is obtained by taking the weighted average of the updated estimates of projections.
The algorithm parameter $\rho$ plays a role in adjusting convergence rate.

\begin{algorithm}[H]
 \caption{Mann iteration for computing PMACE solution}
 \label{algorithm_PMACE}
 \begin{algorithmic}[1]
 \renewcommand{\algorithmicrequire}{\textbf{Input:}}
 \renewcommand{\algorithmicensure}{\textbf{Output:}}
 \REQUIRE Initialization: $x^{(0)} \in \mathbb C^{{N_1} \times N_{2}}$
 \ENSURE  Final Reconstruction: $\hat{x} \in \mathbb C^{{N_1} \times N_{2}}$
 \\ \STATE $\mathbf{w} = \mathbf{v} = [x^{(0)}_{0},\dots, x^{(0)}_{J-1}]^t,\; \text{where} \; x^{(0)}_{j} = P_{j} {x}^{(0)} $ 
  \WHILE {not converged}
  \STATE $\mathbf{w} \gets \mathbf{F} ( \mathbf{v} )$
  \STATE $\mathbf{z} \gets \mathbf{G} (2\mathbf{w} - \mathbf{v})$
  \STATE $\mathbf{v} \gets \mathbf{v} + 2 \rho (\mathbf{z} - \mathbf{w})$
  \ENDWHILE
 \RETURN $\hat{x} = \Lambda^{-1} \sum_{j=0}^{J-1} P_{j}^{t} |D|^{\kappa} v_{j}$ 
 \end{algorithmic} 
\end{algorithm}

\subsection{SHARP+ Algorithm}
\label{sec:sharp+}
The SHARP algorithm \cite{marchesini2016sharp} is an implementation of the RAAR phase retrieval algorithm for ptychography. It operates on images formed by probe illumination; each image is called a frame, defined as $s_{j} = D P_{j} x$, where $P_j$ is the same projection operator defined above.  SHARP defines an operator $P_a$ that ensures the frame data $s_j$ matches the phaseless measurement $y_j$ and an operator $P_Q$ that ensures the overlapping frame data is consistent. These operators are given by stacking the per-patch operators defined by
\begin{align}
    P_{a,j} (s_{j}) &= \mathcal{F}^* \left (  y_{j} \frac{\mathcal{F} s_{j}}{| \mathcal{F} s_{j} |}\right ) \label{eq:sharp1} \\
    P_{Q,j} (s_{j}) &= D P_{j} \left ( \sum_k P_{k}^t |D|^2 \right )^{-1} \sum_i P_{i}^t D^{*} s_{i} \;.\label{eq:sharp2}
\end{align}

SHARP uses the update rule in \cite{luke2004relaxed} with the operators $P_{a}$ and $P_{Q}$ to update the estimate of $s_j$. However, the implementation in \cite{marchesini2016sharp} has a sign change relative to \cite{luke2004relaxed}. In SHARP+, we use the update from \cite{luke2004relaxed}, as in line 3 of Algorithm \ref{algorithm_SHARP_plus}.

Note that the SHARP+ operators \eqref{eq:sharp1} and \eqref{eq:sharp2} are similar to the PMACE operators \eqref{eq:F-interpolate} and \eqref{eq:xjbar}, respectively.  Some differences are (i) PMACE updates operate on image patches rather than frames, which are image patches times probe; (ii) the averaging in \eqref{eq:F-interpolate} takes the place of a relaxation in the update rule of SHARP+ ($\beta$ in Algorithm~\ref{algorithm_SHARP_plus}); (iii) the weighted averaging in \eqref{eq:xjbar} is tuned to match data statistics rather than fixed as the square of the probe as in \eqref{eq:sharp2}.

\begin{algorithm}[H]
 \caption{SHARP+ algorithm}
 \label{algorithm_SHARP_plus}
 \begin{algorithmic}[1]
 \renewcommand{\algorithmicrequire}{\textbf{Input:}}
 \renewcommand{\algorithmicensure}{\textbf{Output:}}
 \REQUIRE Initialization: $\mathbf{x}^{(0)} \in \mathbb C^{{N_1} \times N_{2}}$
 \ENSURE  Final Reconstruction: $\hat{x} \in \mathbb C^{{N_1} \times N_{2}}$
 \\ \STATE $\mathbf{s} = [s_{0},\dots, s_{J-1}]^t , \text{ where } s_{j} = D P_{j} {x}^{(0)} $ 
  \WHILE {not converged}
  \STATE $\mathbf{s} \gets \left [ 2 \beta P_{Q} P_{a} +(1 - 2 \beta) P_{a} - \beta ( P_{Q} - I)  \right ] \mathbf{s}$
  \ENDWHILE
 \RETURN $\hat{x} = \left ( \sum_{k} P_{k}^t |D|^2 \right )^{-1} \sum_{j} P_{j}^t D^{*} s_{j}$
 \end{algorithmic}
\end{algorithm}

\section{Experimental Results}
\label{sec:results}
In this section, we explain our data simulation method and show reconstruction results for synthetic data using PMACE, SHARP+, SHARP, and accelerated Wirtinger flow (AWF).

\begin{figure}[ht]
    \vspace{-0.4cm}
    \centering
    \begin{minipage}[b]{.24\linewidth}
        \centering
        \centerline{\includegraphics[width=2.5cm]{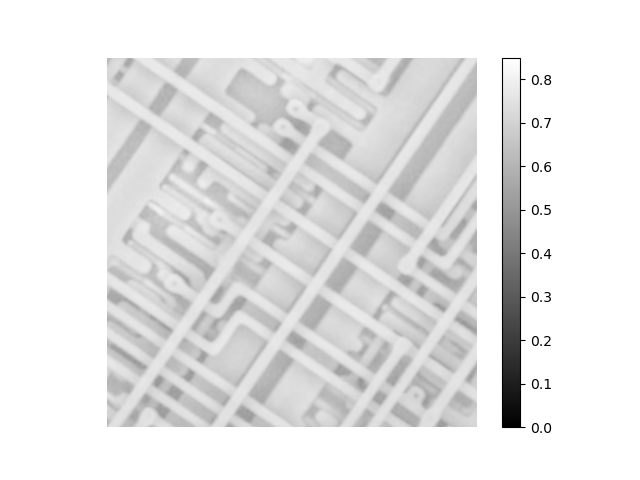}}
       \vspace{-0.2cm}
        \centerline{(a)}
    \end{minipage}
    \begin{minipage}[b]{.24\linewidth}
        \centering
        \centerline{\includegraphics[width=2.5cm]{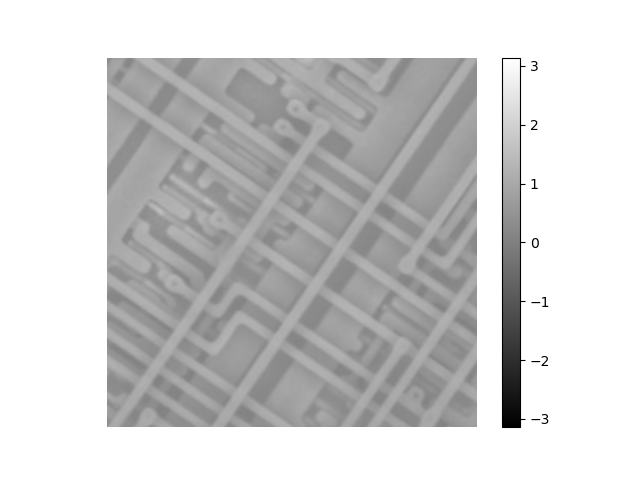}}
       \vspace{-0.2cm}
        \centerline{(b)}
    \end{minipage}
    \begin{minipage}[b]{.24\linewidth}
        \centering
        \centerline{\includegraphics[width=2.5cm]{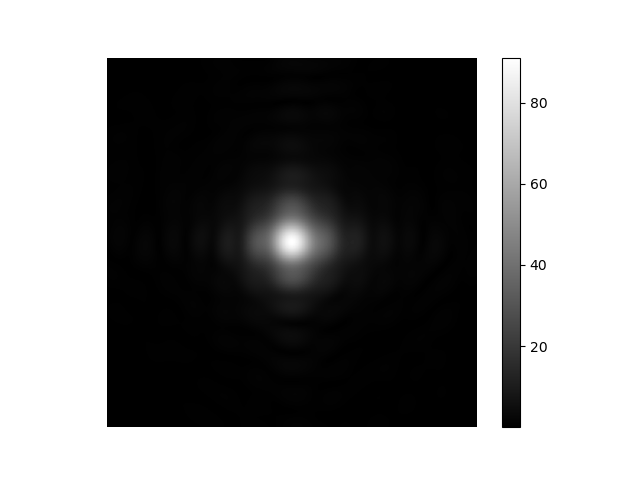}}
        \vspace{-0.2cm}
        \centerline{(c)}
    \end{minipage}
    \begin{minipage}[b]{.24\linewidth}
        \centering
        \centerline{\includegraphics[width=2.5cm]{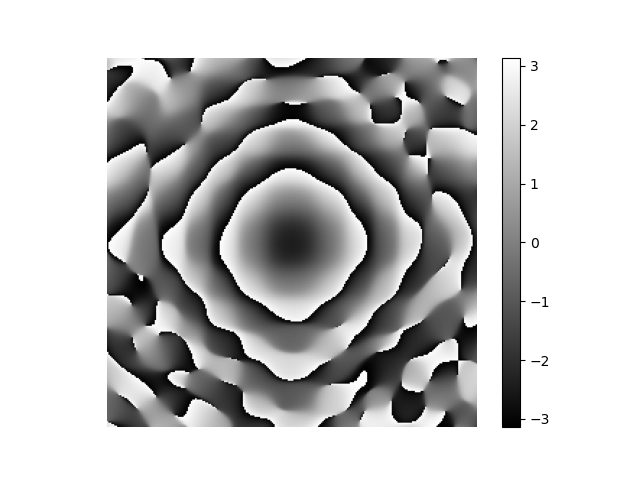}}
        \vspace{-0.2cm}
        \centerline{(d)}
    \end{minipage}
    \vspace{-0.6cm}
    \caption{Complex image and probe used for generating synthetic data. (a) magnitude and (b) phase of ground truth image; (c) magnitude and (d) phase of complex probe.}
    \label{fig:ground_truth_img}
\end{figure}

\begin{figure}[h]
    \centering
    \vspace{-0.2cm}
    \includegraphics[width=3cm]{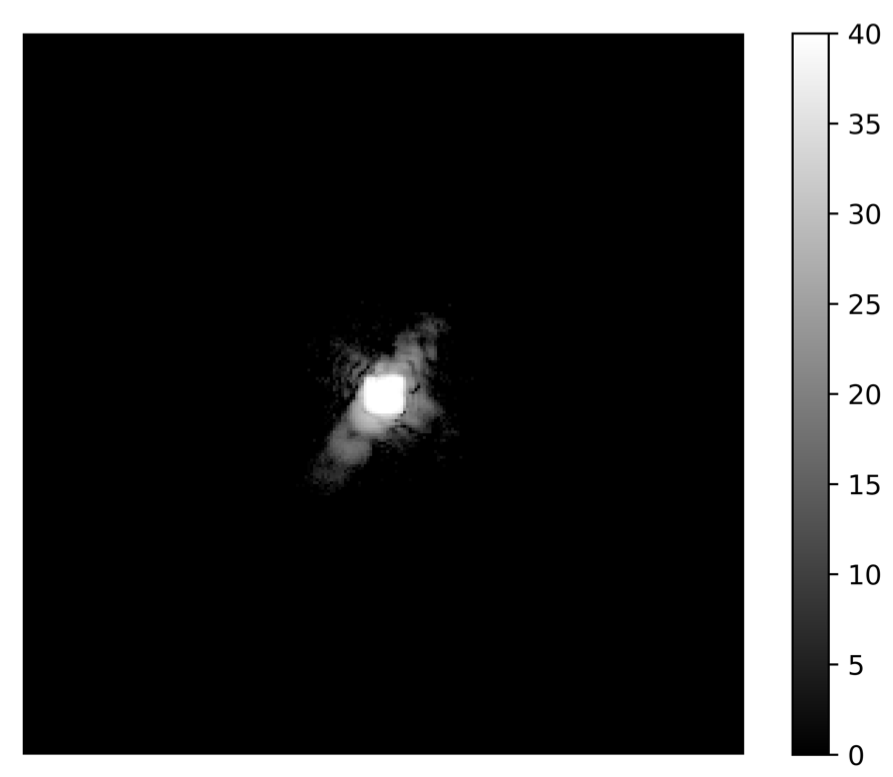}
    \vspace{-0.2cm}
    \caption{Noisy diffraction pattern for reconstruction in dB.}
    \label{fig:diffr_img}
    \vspace{-0.2cm}
\end{figure}

We generate data using the test images shown in Figure \ref{fig:ground_truth_img}. 
The complex transmittance image has size $660 \times 660$ pixels, and the complex probe function has size $256 \times 256$ pixels. 
The probe locations form an $8 \times 8$ grid with all side lengths equal to 56 pixels; this grid is centered inside the full image.  
To simulate the measured diffraction pattern, we extract the projection of the transmittance image at each probe location and multiply by the complex probe function. 
Then we compute the 2D Discrete Fourier Transform to obtain the noise-free synthetic data.
For the noisy synthetic case, we let $r_{p}$ denote the peak photon rate and use this to scale the mean of a Poisson distribution to obtain simulated measurements
\begin{equation}
\label{eq: poisson_rand_generator}
    \hat{y}_{j} \gets \sqrt{\mathrm{Pois}\left(\frac{|\mathcal{F} D x_{j}|^2}{\mathrm{max}(|\mathcal{F} D x_{j}|^{2})} \times r_{p} \right)} \;,
\end{equation}
where $\mathrm{max}(\cdot)$ denotes taking the maximum value of its argument. As $r_{p}$ increases, the signal-to-noise ratio also increases. We take $r_{p}=10^5$ for our simulated noisy diffraction patterns. 
Figure \ref{fig:diffr_img} shows the noisy synthetic data in decibels.

We use Normalized Root-Mean-Square Error (NRMSE) to evaluate the reconstruction results. Since the measured data is independent of a constant phase shift in the full transmittance image, we incorporate this phase shift in NRMSE between the reconstructed complex image $\hat{x}$ and the ground truth image $x$ to obtain 
\begin{equation}
    \begin{aligned}
         e &= \frac{\| \hat{x}- e^{i \theta} x \|}{\| x \|} \;,
    \end{aligned}
\end{equation}
where $\theta \in [0, 2\pi)$ is chosen to minimize the numerator. 

We apply PMACE with fixed probe exponent $\kappa = 1.25$ and Mann averaging parameter $\rho=0.5$. We do a grid search to optimize the noise-to-signal parameter, $\alpha$, of PMACE and the algorithmic parameters of AWF, SHARP,  and SHARP+ algorithms. Then we compare the reconstruction results after 100 iterations of each algorithm. 

\begin{figure}[ht]
  \vspace{-0.4cm}
  \centering
  \begin{minipage}[b]{.24\linewidth}
    \centering
    \centerline{\includegraphics[width=2.5cm]{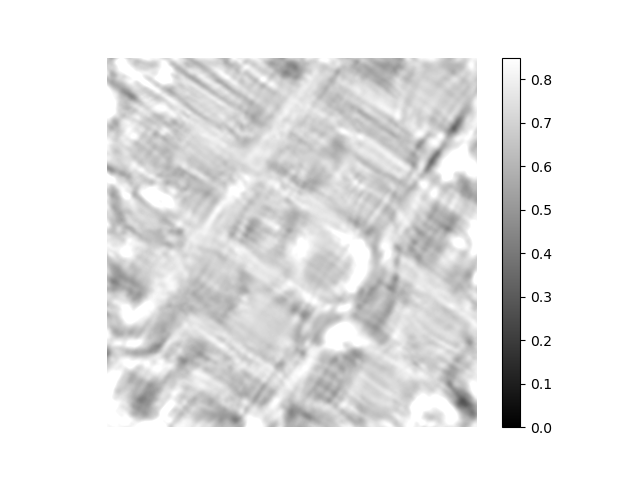}}
    \vspace{-0.1cm}
    \centerline{(a) AWF} 
  \end{minipage}
  \begin{minipage}[b]{.24\linewidth}
    \centering
    \centerline{\includegraphics[width=2.5cm]{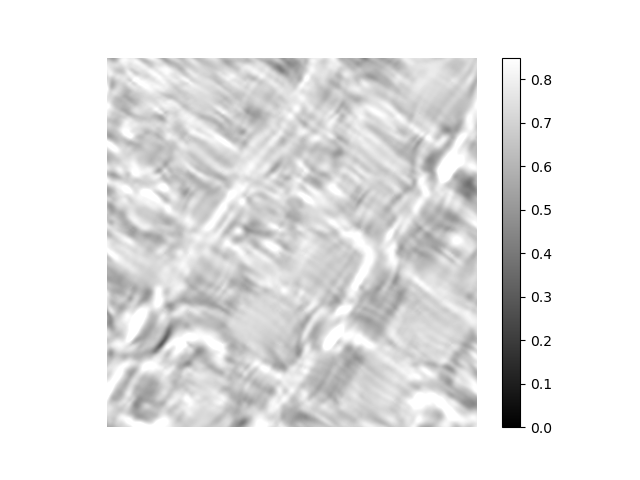}}
    \vspace{-0.1cm}
    \centerline{(b) SHARP} 
  \end{minipage}
  \begin{minipage}[b]{.24\linewidth}
    \centering
    \centerline{\includegraphics[width=2.5cm]{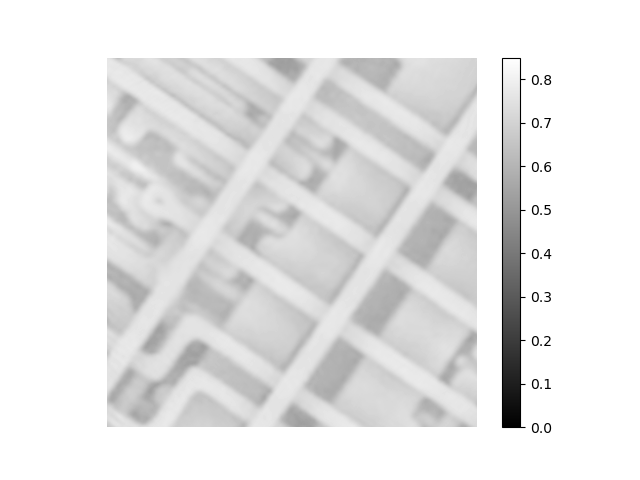}}
    \vspace{-0.1cm}
    \centerline{(c) SHARP+} 
  \end{minipage}
  \begin{minipage}[b]{.24\linewidth}
    \centering
    \centerline{\includegraphics[width=2.5cm]{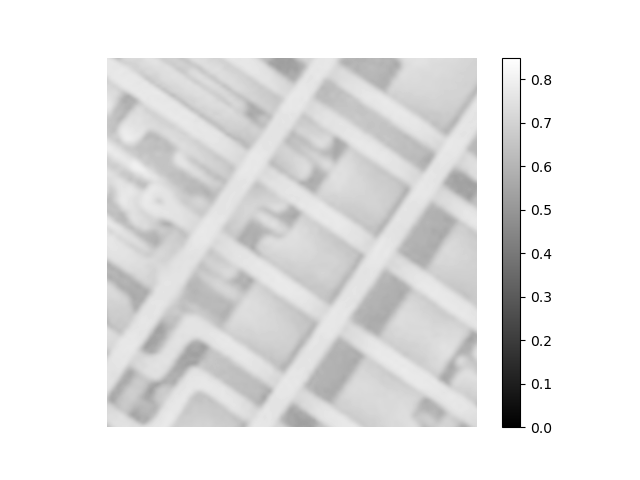}}
    \vspace{-0.1cm}
    \centerline{(d) PMACE} 
  \end{minipage}
  \begin{minipage}[b]{.24\linewidth}
    \centering
    \centerline{\includegraphics[width=2.5cm]{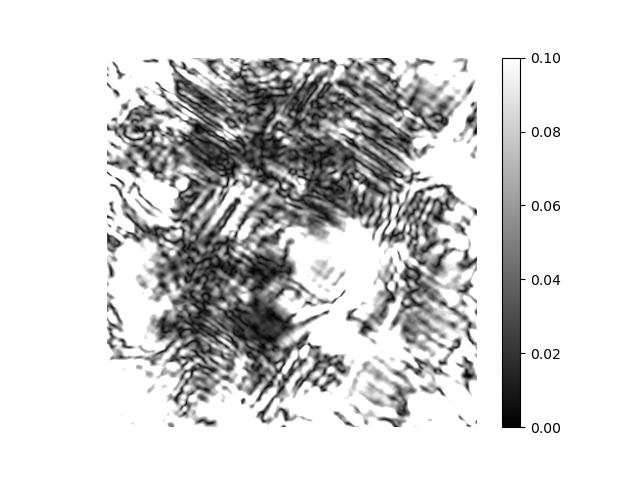}}
    \vspace{-0.1cm}
    \centerline{(e) $e$=0.1586}  
  \end{minipage}
  \begin{minipage}[b]{.24\linewidth}
    \centering
    \centerline{\includegraphics[width=2.5cm]{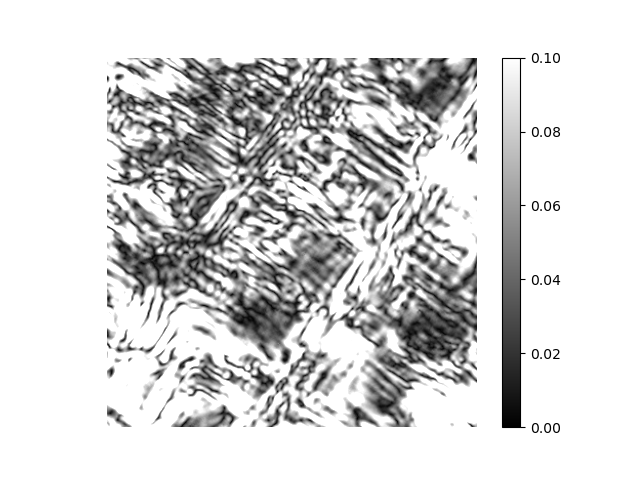}}
    \vspace{-0.1cm}
    \centerline{(f) $e$=0.1445} 
  \end{minipage}
  \begin{minipage}[b]{.24\linewidth}
    \centering
    \centerline{\includegraphics[width=2.5cm]{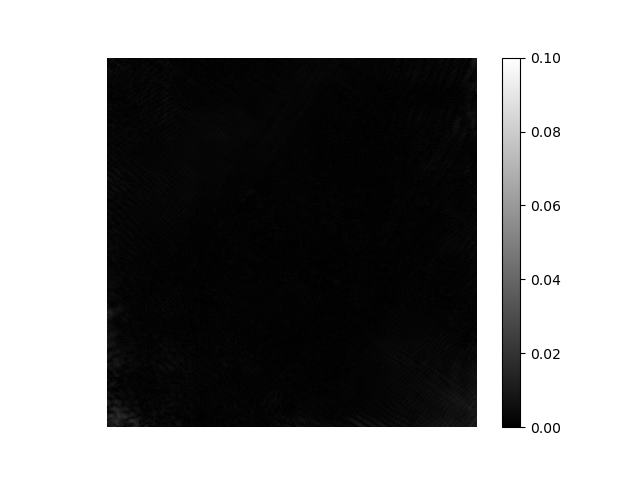}}
    \vspace{-0.1cm}
    \centerline{(g) $e$=0.0033} 
  \end{minipage}
  \begin{minipage}[b]{.24\linewidth}
    \centering
    \centerline{\includegraphics[width=2.5cm]{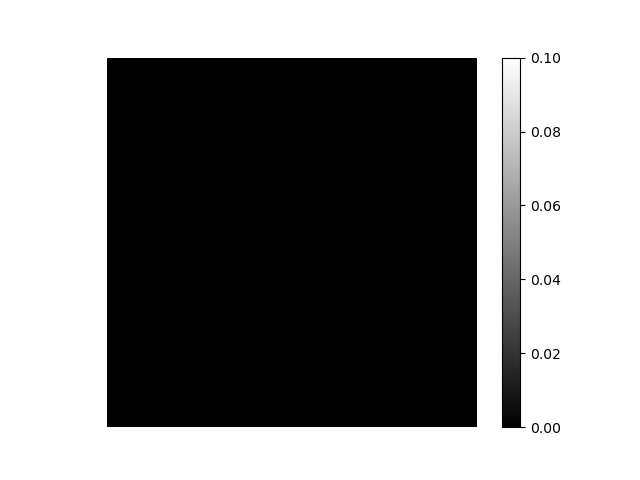}}
    \vspace{-0.1cm}
    \centerline{(h) $e$=0.0002} 
  \end{minipage}
  \vspace{-0.6cm}
  \caption{Reconstructed amplitudes from noiseless data. Top row shows the amplitude images of AWF, SHARP, SHARP+, and PMACE reconstructions. Bottom row shows amplitude of the difference between each complex reconstruction and ground truth along with the final NRMSE.}
  \label{fig:noiseless_results_mag}
  \vspace{-0.3cm}
\end{figure}

\begin{figure}[ht]
  \vspace{-0.3cm}
  \centering
  \begin{minipage}[b]{.24\linewidth}
    \centering
    \centerline{\includegraphics[width=2.5cm]{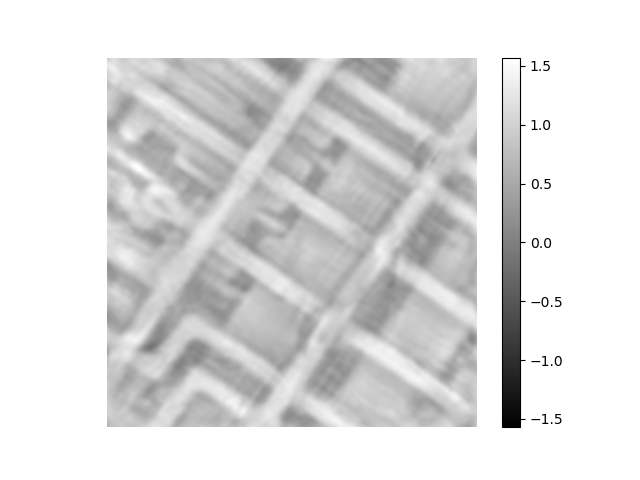}}
    \vspace{-0.1cm}
    \centerline{(a) AWF} 
  \end{minipage}
  \begin{minipage}[b]{.24\linewidth}
    \centering
    \centerline{\includegraphics[width=2.5cm]{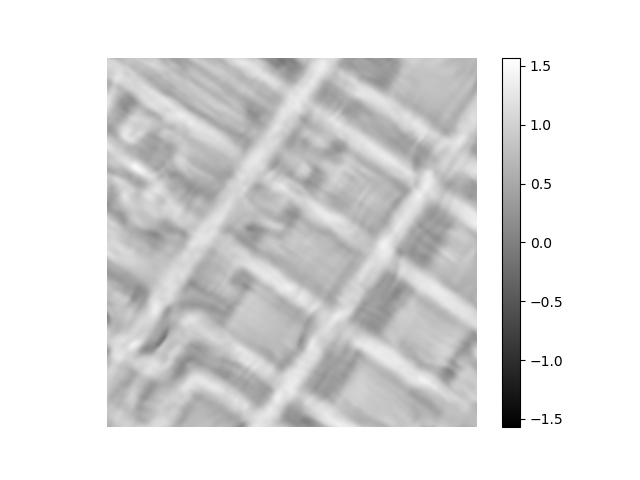}}
    \vspace{-0.1cm}
    \centerline{(b) SHARP} 
  \end{minipage}
  \begin{minipage}[b]{.24\linewidth}
    \centering
    \centerline{\includegraphics[width=2.5cm]{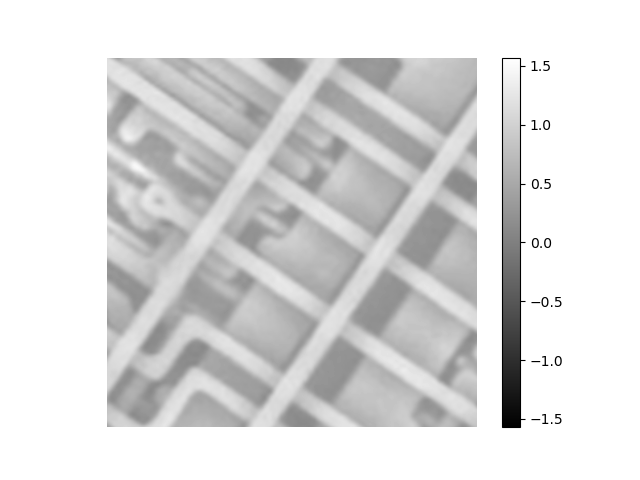}}
    \vspace{-0.1cm}
    \centerline{(c) SHARP+} 
  \end{minipage}
  \begin{minipage}[b]{.24\linewidth}
    \centering
    \centerline{\includegraphics[width=2.5cm]{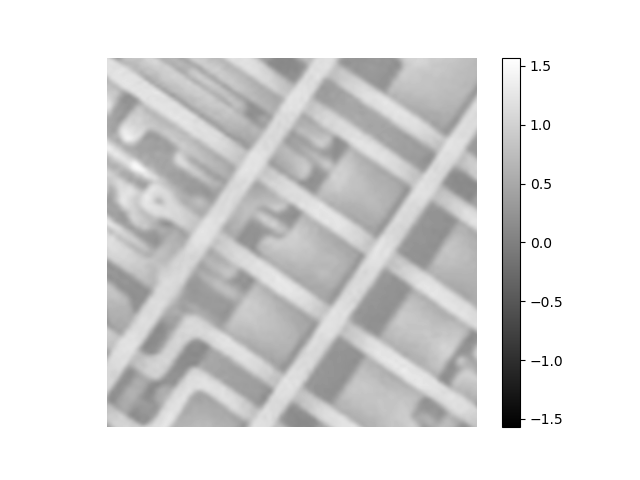}}
    \vspace{-0.1cm}
    \centerline{(d) PMACE} 
  \end{minipage}
  \begin{minipage}[b]{.24\linewidth}
    \centering
    \centerline{\includegraphics[width=2.5cm]{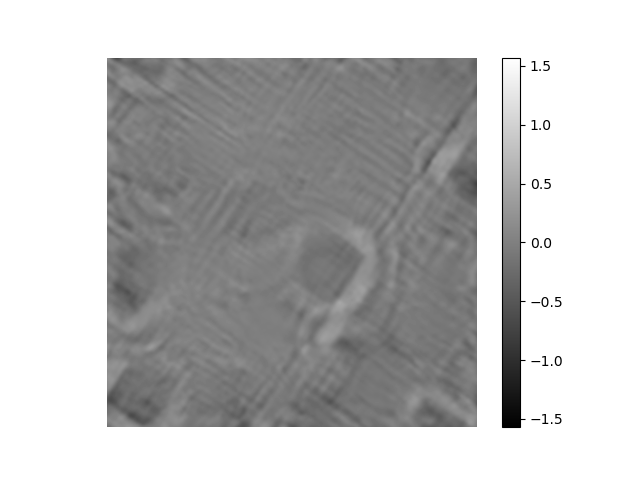}}
    \vspace{-0.1cm}
    \centerline{(e) $e$=0.1586} 
  \end{minipage}
  \begin{minipage}[b]{.24\linewidth}
    \centering
    \centerline{\includegraphics[width=2.5cm]{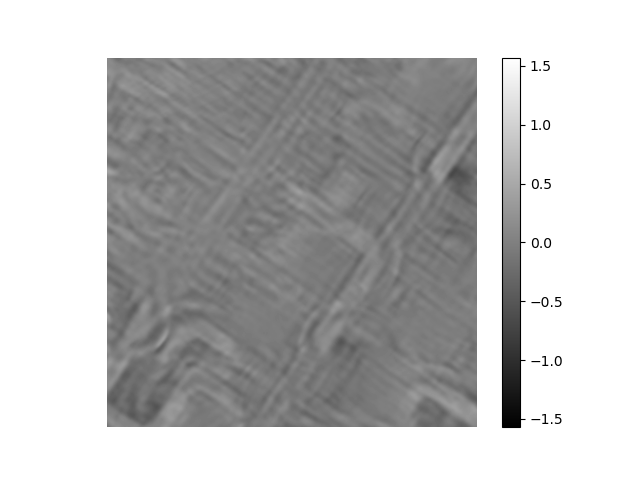}}
    \vspace{-0.1cm}
    \centerline{(f) $e$=0.1445} 
  \end{minipage}
  \begin{minipage}[b]{.24\linewidth}
    \centering
    \centerline{\includegraphics[width=2.5cm]{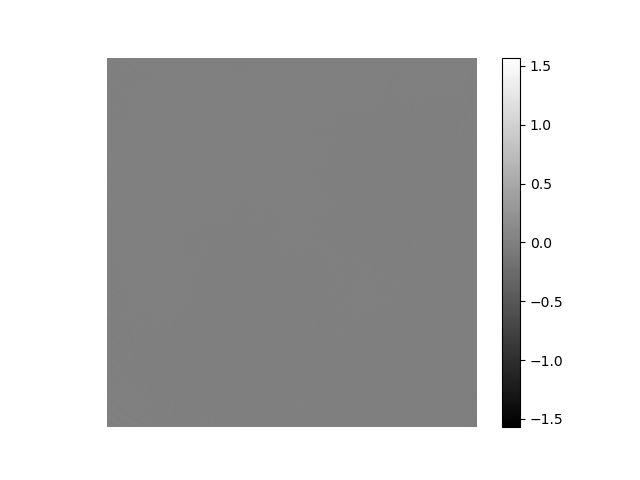}}
    \vspace{-0.1cm}
    \centerline{(g) $e$=0.0033} 
  \end{minipage}
  \begin{minipage}[b]{.24\linewidth}
    \centering
    \centerline{\includegraphics[width=2.5cm]{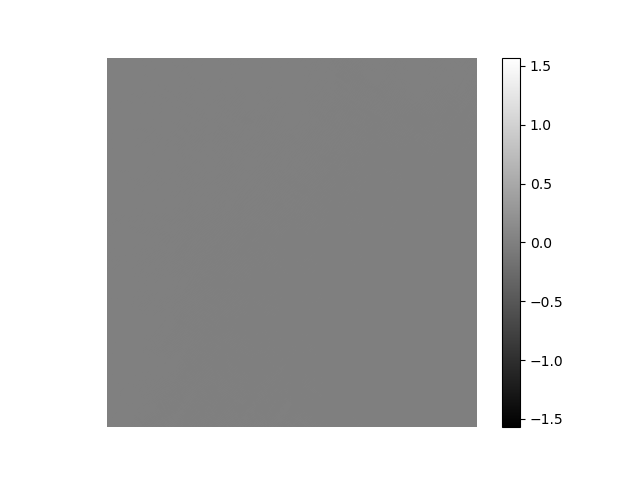}}
    \vspace{-0.1cm}
    \centerline{(h) $e$=0.0002} 
  \end{minipage}
  \vspace{-0.6cm}
  \caption{Reconstructed phases from noiseless data. Top row shows the phase images for AWF, SHARP, SHARP+, and PMACE reconstructions.  Bottom row shows the phase difference between each complex reconstruction and ground truth. }
  \label{fig:noiseless_results_phase}
\end{figure}

\begin{figure}[H]
    \centering
    \vspace{-0.8cm}
    \includegraphics[width=7.6cm]{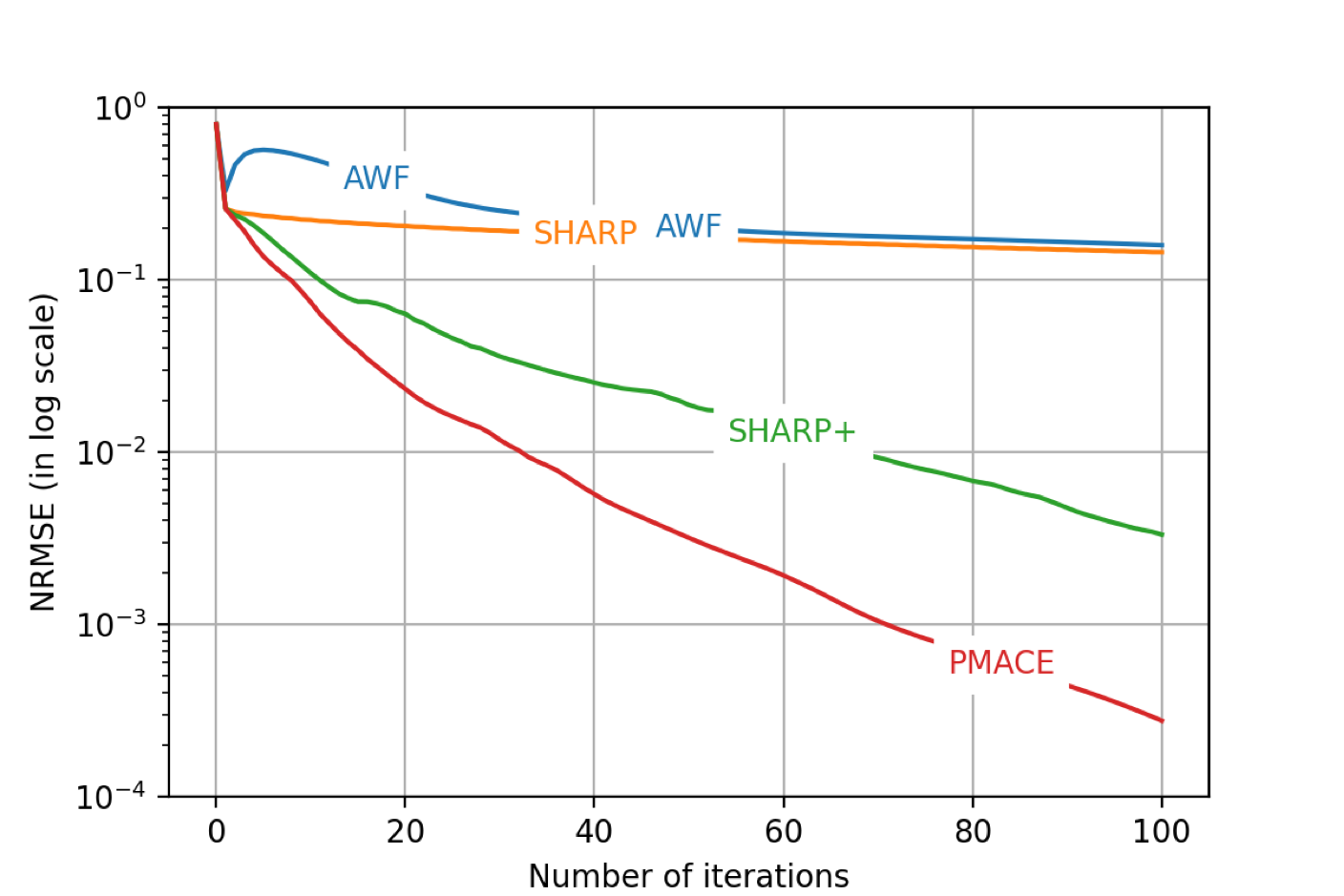}
    \vspace{-0.2cm}
    \caption{Convergence plots for reconstruction on noiseless data.}
    \label{fig:noiseless_convergence_plots}
    \vspace{-0.2cm}
\end{figure}

The upper plots of Figure \ref{fig:noiseless_results_mag} show the reconstructed amplitudes from AWF, SHARP, SHARP+ and PMACE using noise-free data. The bottom plots show the corresponding amplitude errors (the amplitude of the difference between a complex reconstruction and the ground truth image) plus the NRMSE values. 
Figure \ref{fig:noiseless_results_phase} shows the reconstructed phases from each algorithm and the phase errors.
Figure \ref{fig:noiseless_convergence_plots} shows NRMSE as a function of number of iterations in the noise-free case. 

We observe that the PMACE approach exhibits much better image quality when compared with AWF and SHARP and somewhat better quality than SHARP+. The convergence plots in Figure \ref{fig:noiseless_convergence_plots} show that PMACE  has very fast convergence relative to these other methods.

\begin{figure}[ht]
    \vspace{-0.1cm}
  \centering
  \begin{minipage}[b]{.24\linewidth}
    \centering
    \centerline{\includegraphics[width=2.5cm]{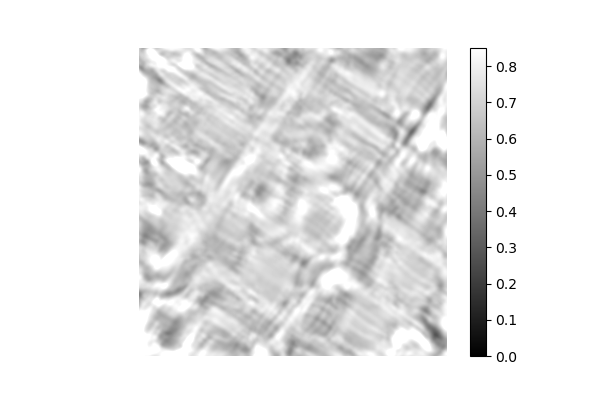}}
    \vspace{-0.1cm}
    \centerline{(a) AWF} 
  \end{minipage}
  \begin{minipage}[b]{.24\linewidth}
    \centering
    \centerline{\includegraphics[width=2.5cm]{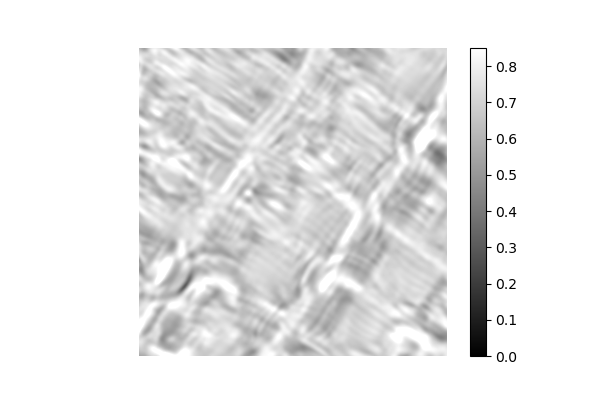}}
    \vspace{-0.1cm}
    \centerline{(b) SHARP} 
  \end{minipage}
  \begin{minipage}[b]{.24\linewidth}
    \centering
    \centerline{\includegraphics[width=2.5cm]{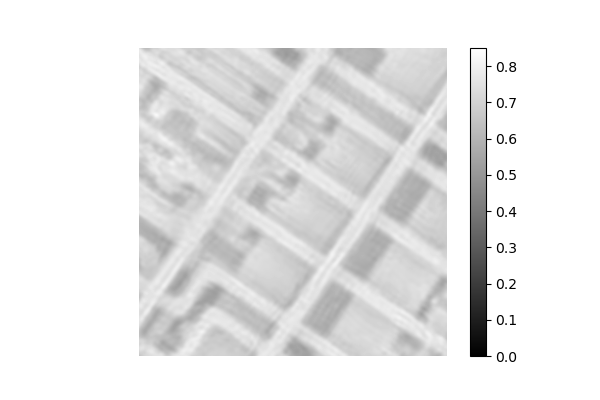}}
    \vspace{-0.1cm}
    \centerline{(c) SHARP+} 
  \end{minipage}
  \begin{minipage}[b]{.24\linewidth}
    \centering
    \centerline{\includegraphics[width=2.5cm]{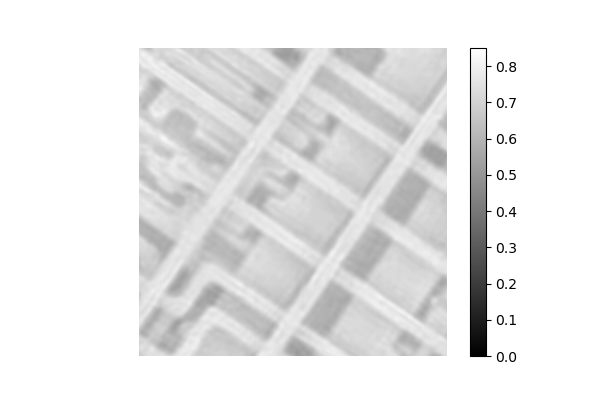}}
    \vspace{-0.1cm}
    \centerline{(d) PMACE} 
  \end{minipage}
  \begin{minipage}[b]{.24\linewidth}
    \centering
    \centerline{\includegraphics[width=2.5cm]{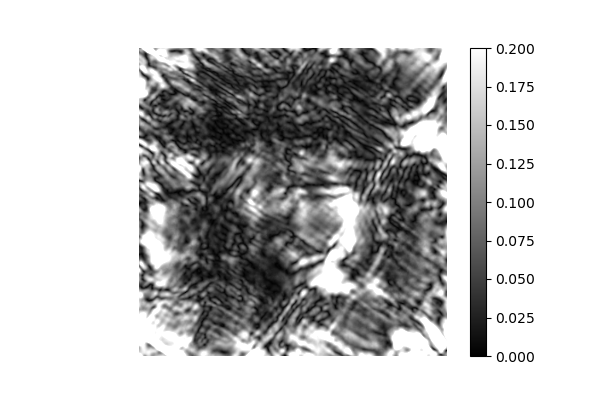}}
    \vspace{-0.1cm}
    \centerline{(e) $e$=0.1655} 
  \end{minipage}
  \begin{minipage}[b]{.24\linewidth}
    \centering
    \centerline{\includegraphics[width=2.5cm]{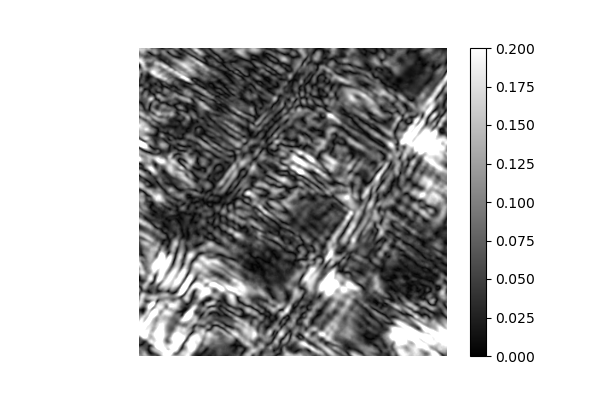}}
    \vspace{-0.1cm}
    \centerline{(f) $e$=0.1385} 
  \end{minipage}
  \begin{minipage}[b]{.24\linewidth}
    \centering
    \centerline{\includegraphics[width=2.5cm]{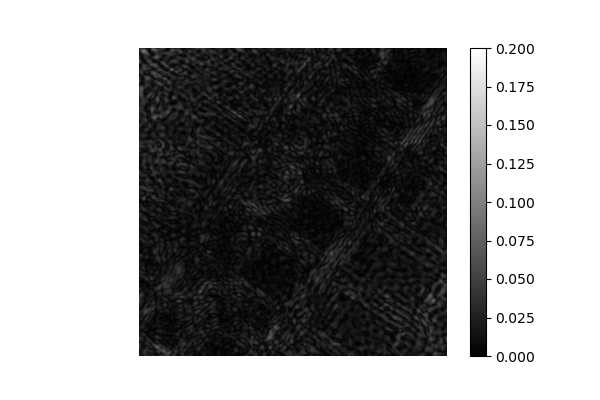}}
    \vspace{-0.1cm}
    \centerline{(g) $e$=0.0272} 
  \end{minipage}
  \begin{minipage}[b]{.24\linewidth}
    \centering
    \centerline{\includegraphics[width=2.5cm]{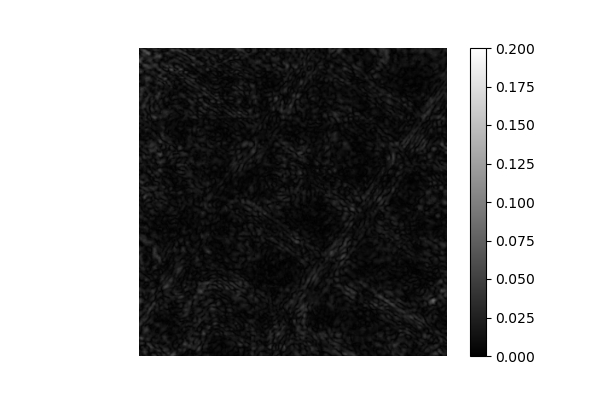}}
    \vspace{-0.1cm}
    \centerline{(h) $e$=0.0210} 
  \end{minipage}
  \vspace{-0.6cm}
  \caption{Reconstructed amplitudes from noisy data. Top row shows the amplitude images for AWF, SHARP, SHARP+, and PMACE reconstructions. Bottom row shows amplitude of the difference between each complex reconstruction and ground truth along with the final NRMSE.}
  \label{fig:noisy_results_mag}
  \vspace{-0.4cm}
\end{figure}

\begin{figure}[ht]
  \centering
  \begin{minipage}[b]{.24\linewidth}
    \centering
    \centerline{\includegraphics[width=2.5cm]{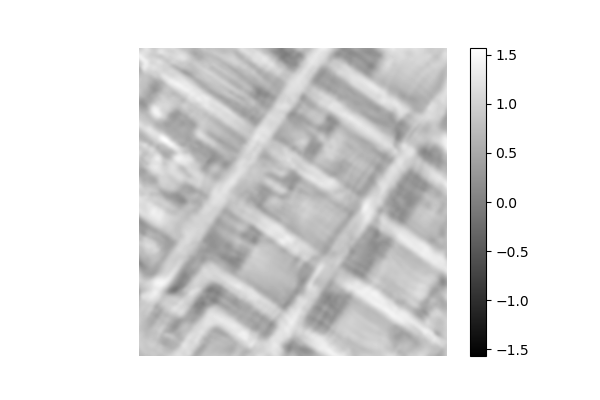}}
    \vspace{-0.1cm}
    \centerline{(a) AWF} 
  \end{minipage}
  \begin{minipage}[b]{.24\linewidth}
    \centering
    \centerline{\includegraphics[width=2.5cm]{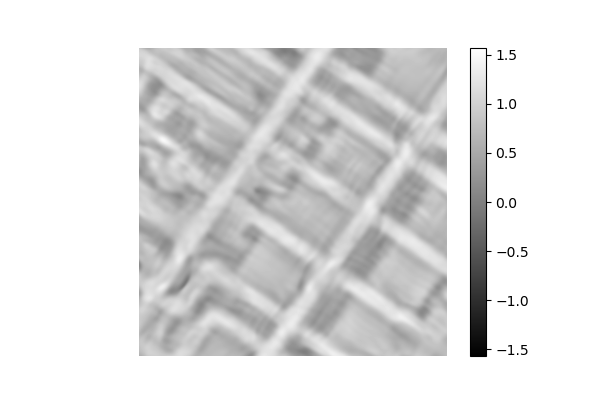}}
    \vspace{-0.1cm}
    \centerline{(b) SHARP}   
  \end{minipage}
  \begin{minipage}[b]{.24\linewidth}
  \centering
    \centerline{\includegraphics[width=2.5cm]{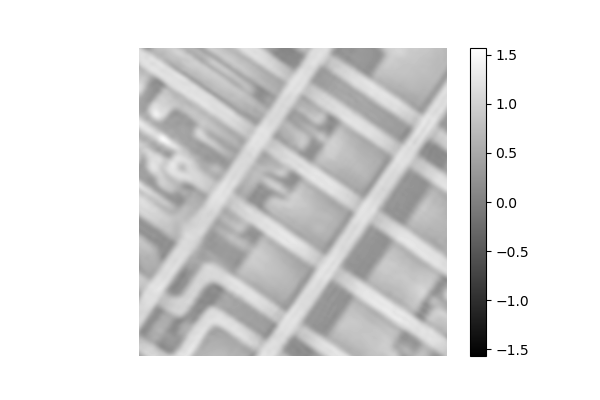}}
    \vspace{-0.1cm}
    \centerline{(c) SHARP+}   
  \end{minipage}
  \begin{minipage}[b]{.24\linewidth}
    \centering
    \centerline{\includegraphics[width=2.5cm]{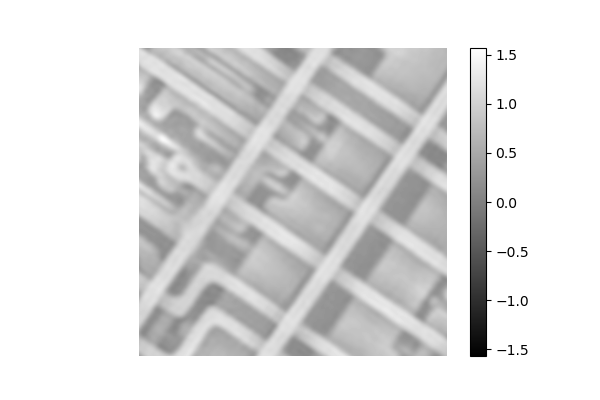}}
    \vspace{-0.1cm}
    \centerline{(d) PMACE}   
  \end{minipage}
  \begin{minipage}[b]{.24\linewidth}
    \centering
    \centerline{\includegraphics[width=2.5cm]{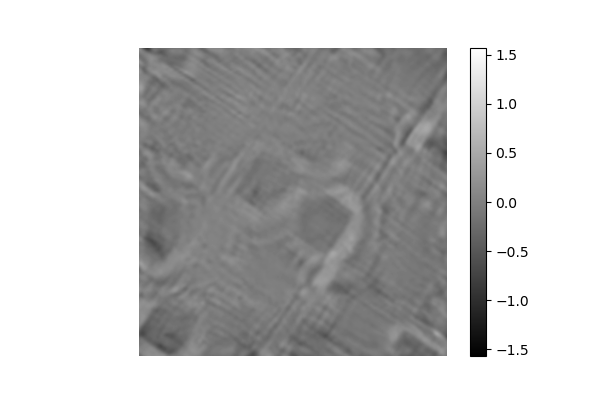}}
    \vspace{-0.1cm}
    \centerline{(e) $e$=0.1655}   
  \end{minipage}
  \begin{minipage}[b]{.24\linewidth}
    \centering
    \centerline{\includegraphics[width=2.5cm]{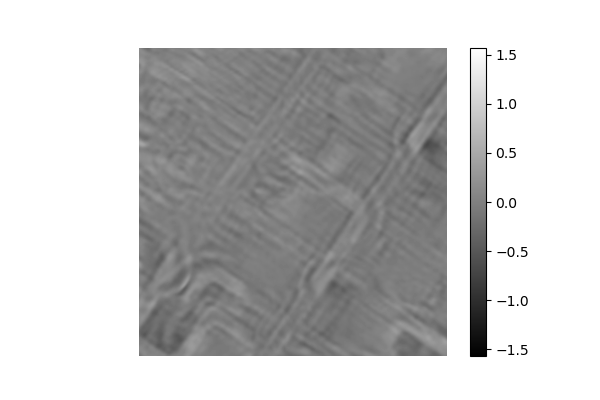}}
    \vspace{-0.1cm}
    \centerline{(f) $e$=0.1385}   
  \end{minipage}
  \begin{minipage}[b]{.24\linewidth}
    \centering
    \centerline{\includegraphics[width=2.5cm]{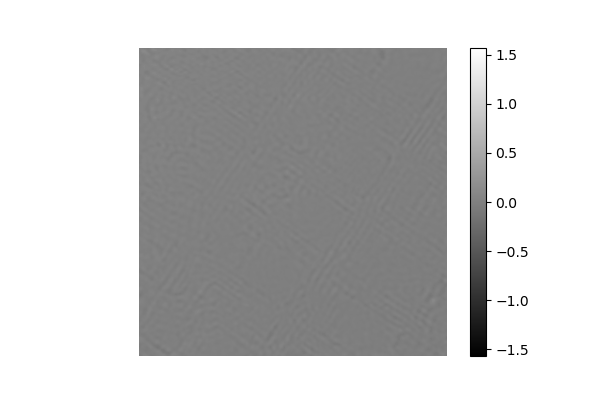}}
    \vspace{-0.1cm}
    \centerline{(g) $e$=0.0272}   
  \end{minipage}
  \begin{minipage}[b]{.24\linewidth}
    \centering
    \centerline{\includegraphics[width=2.5cm]{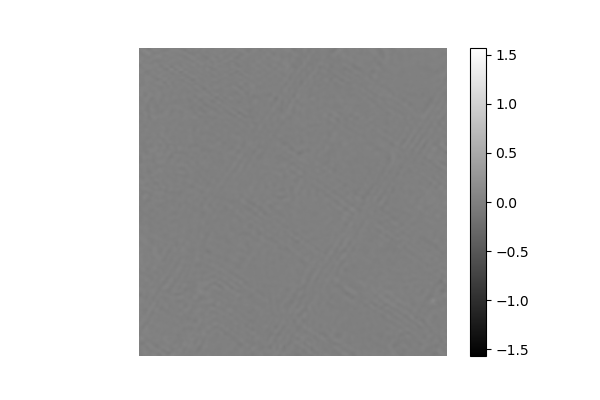}}
    \vspace{-0.1cm}
    \centerline{(h) $e$=0.0210}   
  \end{minipage}
  \vspace{-0.6cm}
  \caption{Reconstructed phases from noisy data. Top row shows the phase images for AWF, SHARP, SHARP+, and PMACE reconstructions. Bottom row shows the phase difference between each complex reconstruction and ground truth.}
  \label{fig:noisy_results_phase}
\end{figure}

\begin{figure}[ht]
    \centering
    \vspace{-0.6cm}
    \includegraphics[width=7.6cm]{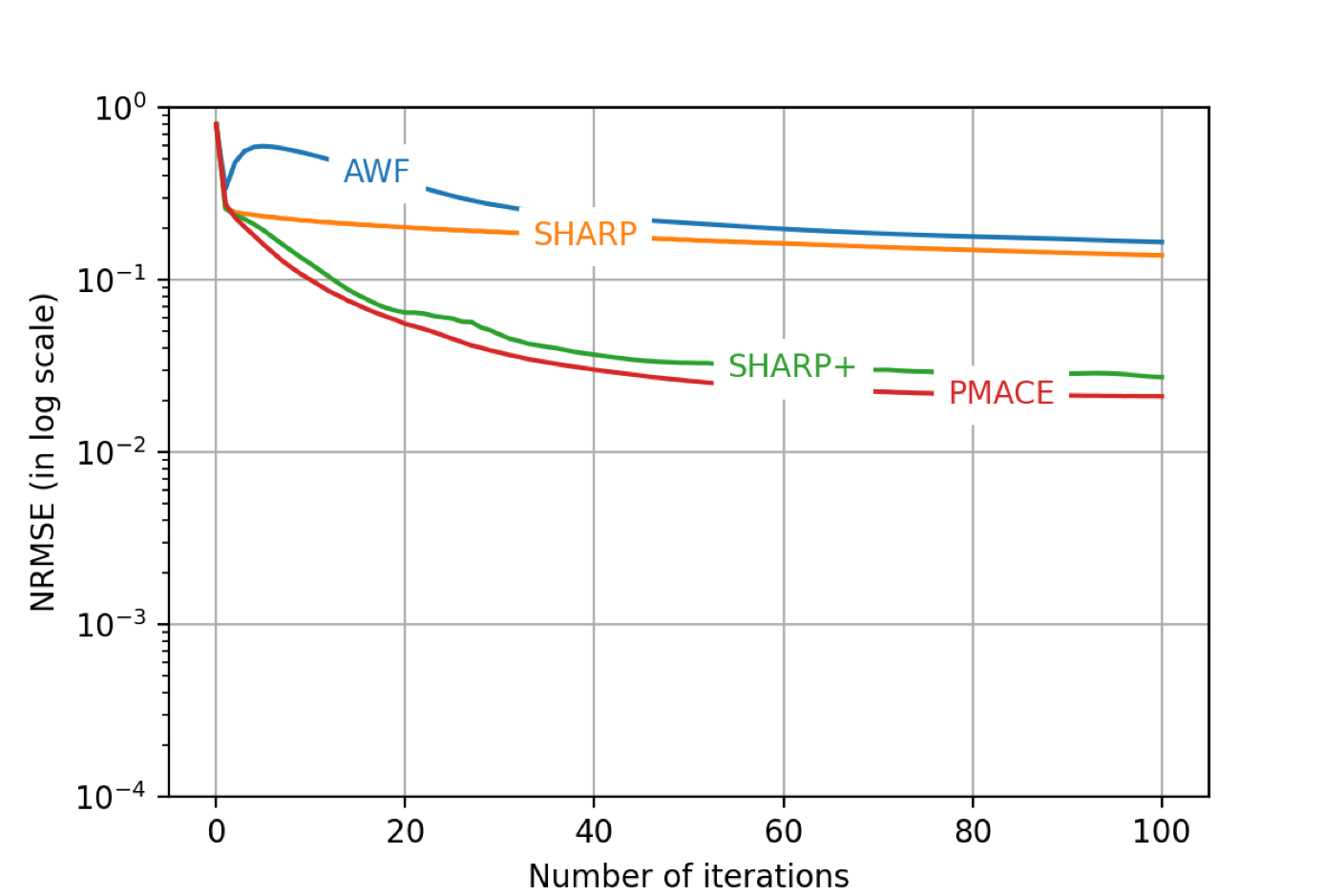}
    \vspace{-0.2cm}
    \caption{Convergence plots for reconstruction on noiseless data.}
    \label{fig:noisy_convergence_plots}
    \vspace{-0.2cm}
\end{figure}

Figures \ref{fig:noisy_results_mag} and \ref{fig:noisy_results_phase} show the corresponding amplitudes and phases of each approach in the noisy synthetic case. Note that the  images reconstructed using PMACE contain the fewest artifacts. The convergence plot for the noisy case, shown in Figure \ref{fig:noisy_convergence_plots}, again shows that PMACE exhibits faster convergence than other algorithms, although in this case, SHARP+ is a close competitor.

\section{Conclusion}
\label{sec:conclusion}
We described the PMACE approach for ptychographic image reconstruction. This approach builds on the MACE framework for describing inverse problems and uses a novel weighting scheme based on the probe amplitude for both the data-fidelity agents and the consensus averaging operator.  PMACE is easily parallelized with guaranteed convergence under appropriate hypotheses.  We also described SHARP+ as a faster-converging variant of the SHARP algorithm.  Simulation results from synthetic data indicate that PMACE outperforms competing algorithms in terms of both convergence speed and reconstruction quality, with SHARP+ a close second in the case of noisy data.

\section*{Acknowledgment}
We thank Dr. Kevin Mertes of Los Alamos National Laboratory for providing the complex image and probe.

\bibliographystyle{IEEEbib}
\bibliography{Asilomar21_PMACE}

\end{document}